# Evaporation induced self-assembling of few layer graphene into fractal-like conductive macro-network: the way to reduce percolation threshold

I. Janowska*

Institut de Chimie et Procédés pour l'Energie, l'Environnement et la Santé(ICPEES), CNRS UMR 7515-University of Strasbourg, 25 rue Becquerel, 67087 Strasbourg (ECPM), France

*e-mail: janowskai@unistra.fr, tel : +33 (0) 3 68 85 26 33

**Abstract**

Transcription of nanoproperties to others dimensions in efficient and simple way by appropriate design of devices is a challenge; and conductive nanocarbon systems are concerned. Here, the evaporation induced self-assembling method is proposed to form branched, fractal-like ''2D'' conductive macrostructures from (nano) micro-few layer graphene flakes. The self-assembled conductive graphene networks reveal a reduction of percolation threshold compared to random arrangement (reduction of matter for a given surface and conduction at lower surface coverage), and becomes a promising matrix for conductive (transparent) films.

Construction of high performance macro(micro) devices in a way that preserve the features of their micro(nano) components still requires the efforts. In 2D or 3D systems (films, polymers, composites) containing conductive nano(micro)carbons, the electrical properties of carbon components are hardly propagated throw the surface or bulk due to high percolation threshold, which besides low carbon loading can be related to an insufficient contact between individual components (morphological and chemical identity dependent interactions) but also to their disordered, random arrangement. Concerning a recent development of (transparent) conductive graphene-based films, the highest qualities are achieved by CVD synthesis and despite the problem of (rotary) grain boundaries between the nano- or macro- sheets low sheet resistance are related to (0.28 -1 kΩ/sq of sheet resistance increasing two order of magnitude for supported graphene).[1-5] The micro(nano-)graphenes originated from top down approach synthesis (exfoliation of graphite based materials) exhibit usually lower conductivity and their (transparent) conductive films formed by spin coating, hot-spray or other techniques show few-few-dozen kΩ sheet resistance.[6-10] Varied transparency/conductivity property of the films is required depend on the final (opto)electronic application (e-readers, electrochromic windows, touch sensors, photovoltaic, etc…);[11] this relationship can be modified by films' thickness within a given matter. A variant is low surface coverage e.g. formation of graphene-carbon nanotubes hybrids, graphene mesh electrodes.[12-15]

At present, we propose the evaporation induced self-assembling method as an alternative way to reach highly conductive graphene (few layer graphene) network at macroscale. The approach leads to branched, fractal-like patterns. In principle, these autoscalable structures achieve the goal of nano→ micro → macro scale multiplication. Apart from branched geometries of trees or river beds widely present in nature, the fractals approach is also present in time or market analysis.[16-19] The formation of self-assembled structures from dried nanofluids is not entirely understood and final morphologies depend on many factors such as solvent, evaporation conditions, particle size, identity, concentration and thermodynamic state, which affect chemical potential, nanoparticles mobility and attractions (repulsions) between particles, particles and substrate, particles and solvent.[20-22] Initially it was described by hydrodynamic instabilities and solute transports phenomena.[23,24] Further, the formation of branched structure could be predicted with Kinetic Monte Carlo (KMC) dynamics and "coarse-grained" two dimensional microscopic lattice gas model based on uniformity of solvent dynamics and the fluctuation of nanoparticles boundaries following evaporation.[25] Following investigations introduced additional instability at the macroscopic solvent-air-substrate contact line in accordance with Marangoni effect.[20,21] The recently reported "open" domain simulations consider also particles aggregation during evaporation process, assisted by shrinking of drying fronts from the edges toward the center.[26]

On the other hand, the aggregation of particles appeared in general as a central problem in the applied colloid science and was described by diffusion-limited aggregation (DLA) and cluster cluster aggregation (CA) mechanisms based on Monte Carlo simulations.[27-31] The aggregation process limited by diffusion of particles to the aggregate surface is similar to precipitation of species from a supersaturated matrix or to a crystal grows from a supercooled melt,[29] while DLA model produces complicated shapes such as dendrits and clusters with fractal dimensionality. Curiously, micro-dendritic and fractals mono and polycrystals were observed for CVD grown graphene islands on copper and gold foils which suggested diffusion-limited process;[32-34] the fractal etching of graphene on liquid copper surface was reported as well.[35] Recently Wang et al. reported on the self-assembly of DNA and biofunctional polymer on graphene nanorribons, where non-covalent interactions (van der Waals, charge transfer) permitted an adhesion of graphene to SiO$_2$ surface and assembling of the macromolecules on graphene.[36,37]

While a formation of branched structures is more elucidated, their impact on percolation phenomenon in the conductive films is rarely studied. The germinal theoretical simulations show that reduction of percolation threshold can occur in self-assembled patterns compared to randomly assembled structures.[38,39]

Two types of few layer graphene (FLG) flakes originated from different synthesis methods are here considered. The FLG obtained by mechanical ablation of pencil lead has an average lateral size of 2.5 µm and multi step structures (the lateral size of sheets differs within a given flake, (FLG-Abl).[40,41] The second FLG was extracted from expanded graphite (FLG-EG) under µ-waves irradiations, it exhibits higher 2D aspect compared to FLG-Abl i.e. few up to dozen of µm lateral sheets size with similar average sheets number (~ 6 and up to 20). Both FLGs have similar atomic oxygen percentage (~5%, XPS), while higher conductivity was measured for FLG-EG. The minimal resistance measured for FLG-EG individual flakes is 500Ω, while for FLG-Abl flakes a min. resistance was of 1.6 kΩ after annealing at 800°-900°C.[40] Three different solvents are used in the evaporation process: chloroform, toluene and ethanol. These are the solvents having an ability to significantly disperse graphene, which aimed to avoid an agglomeration of the flakes prior to evaporation processes. A diffusion rate was additionally influenced by evaporation time; the samples are left to dry in open or closed systems. The ethanol suspension of FLG-Abl was drop casted on flat glass plate and left as such for drying (f-1), two toluene suspension of FLG-Abl were left in the closed glass vessel with two different concentrations (f-2 higher, f-3 lower); and chloroform suspension of FLG-EG was drop casted on flat glass plate and left for evaporation in closed vessel.

Once the evaporation processes finished, all investigated FLGs formed well- organized continuous branched patterns (f-1, f-2 in Fig.1, f-3 in Fig.2 and f-4 in Fig.3), which in majority can be detected by eyes; and mostly with fractal characteristics.

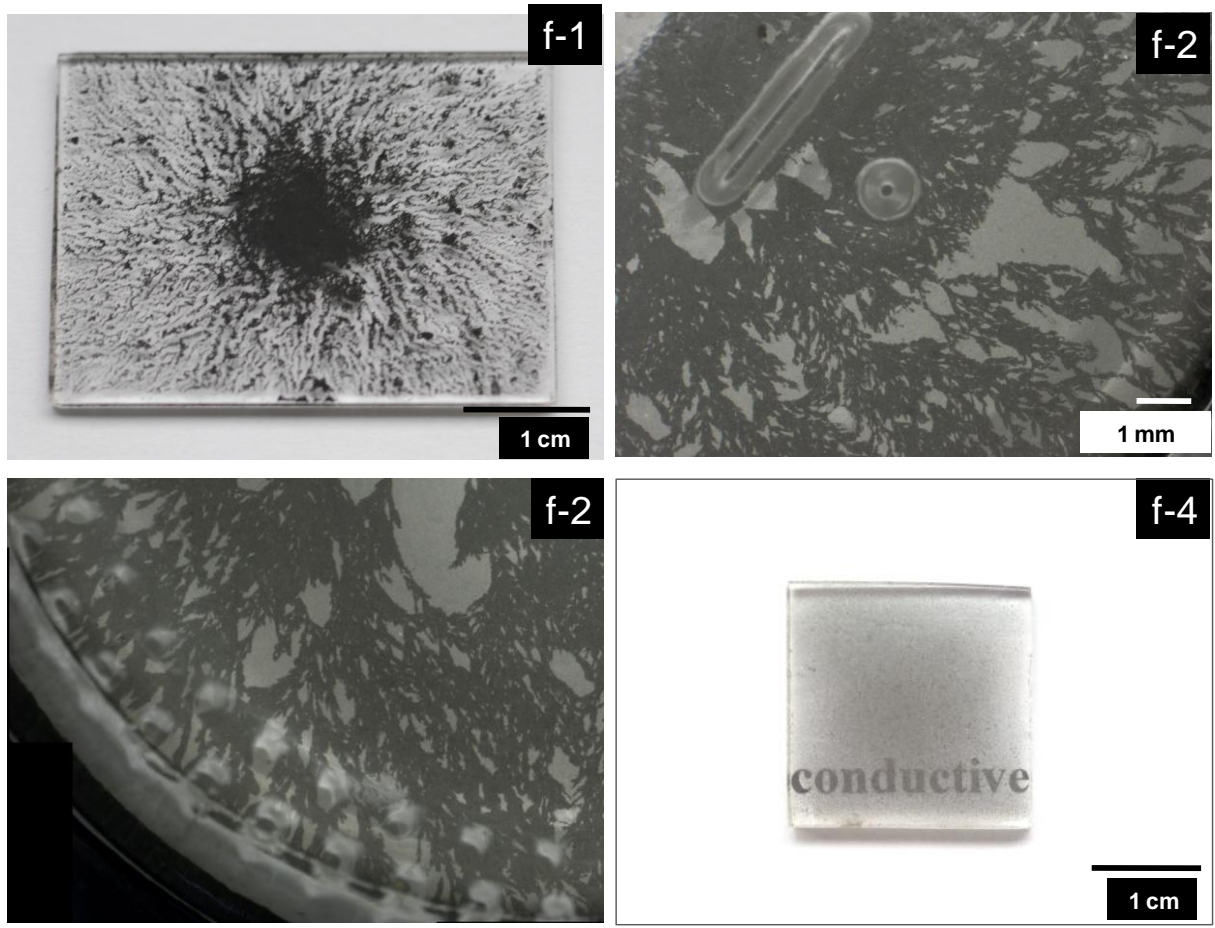

Figure1.Optical photos of self-assembled patterns formed via evaporation: f-1 from FLG-pencil in ethanol in open system on flat glass plate, f-2 from FLG-pencil in toluene in closed system on bossy round substrate, f-4 from FLG-EG in chloroform in closed system on glass plate.

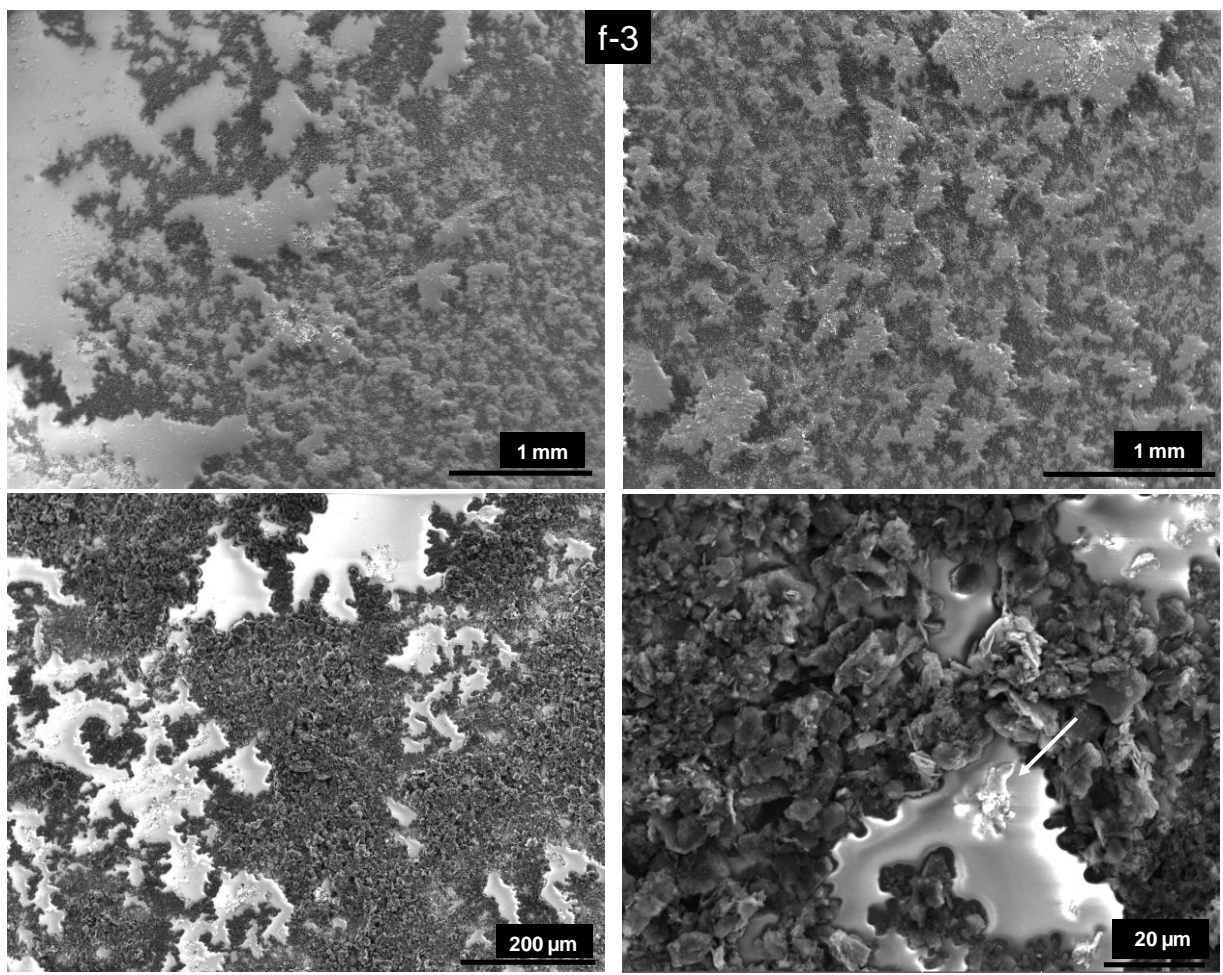

Figure 2. SEM micrographs of self-assembled pattern of FLG-pencil formed via evaporation of toluene in closed system (f-3).

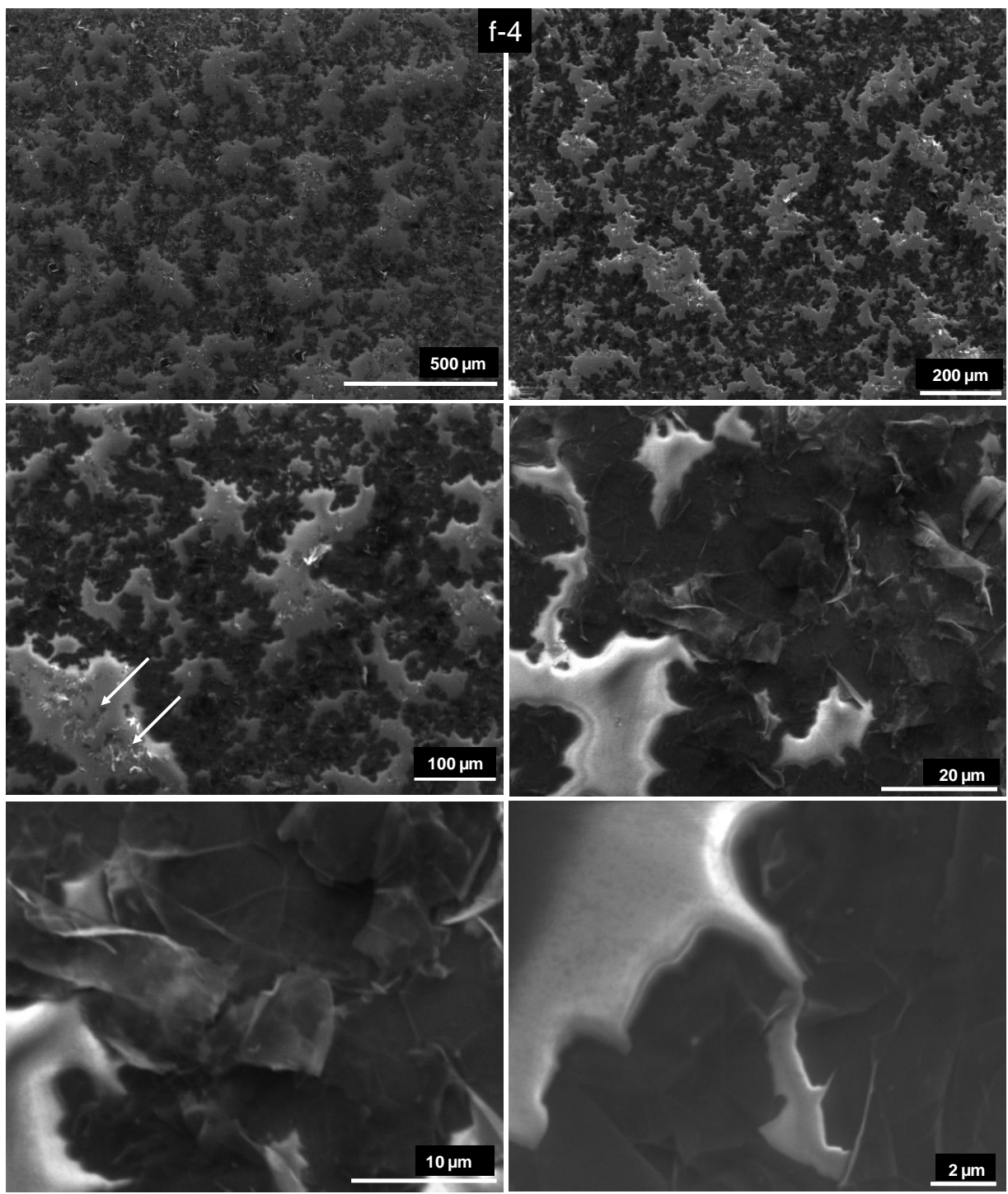

Figure 3. SEM micrographs of self-assembled pattern of expanded graphite originated few layer graphene formed via evaporation of chloroform in closed system on glass plate (f-4).

The observed in literature branched structures are usually µm size, here the continuous networks cover all (or *quasi*-all) centimeters range substrate surfaces revealing that size-unlimited substrates can be potentially used for such grown structures. Affected by evaporation conditions and FLG identity, the morphology of formed patterns varies. The f-1,

f-3 and f-4 can be considered as one integral macroscopic network, f-2 is a network constituted of connected small trees structures. All patterns although have a "2D" aspect (film) revealing a good dispersion of the flakes in the solvents and significant interactions of the flakes with glass substrates. The certain degree of 3D aggregations, i.e. stacking of the flakes in z directions, is related to van der Walls attractions, which are strengthened by quite high initial concentration of the flakes. F-4 self-assembly, formed from FLG-EG lie in flatter manner than FLG-Abl originated patterns (HR-SEM fig.3 versus fig.2), which is in agreement with higher 2D aspect (lateral size/thickness) of its components. In the patterns formed from FLG-Abl, the flakes have tendency to lift from the surface and stack each other at higher angle from the surface, which is due to their lower 2D feature and/or multi step structure.[41] In traditional approach they assimilate rather "particles" than "sheets" behavior. The observed densification of matter in a center of f-1, f-2, and f-3 is in accordance with "open" domain simulations, where an amount of flakes are pulled toward the center after shrinking of drying front.[26] F-4 network exhibits more homogenous overall density and "order" of branches, where main branches initiate from one side of the substrate plate. The main branches have seeds in a center of plate in fractals f-1, and f-3 (fig.1, fig.3) and their density as well as a density of sub-branches decrease with increase of substrate radius from that center. In f-2 (fig.1), where a round, bossy surface and long time evaporation was applied, the overall pattern consists of three connected circular networks of trees which keep the same morphology but their size and (sub-)branches densities change periodically with the substrate radius (density decreases and size of substructures increases with the radius increase). This circular symmetry suggests a uniform evaporation and density of the FLG particles on the overall surface (number of flakes/substrate area), while the substrate center remains FLG-free and the seeds of trees are arranged at three different radius circular lines due to the bossy form of the substrate.

The AFM topography profiles performed for f-4 (fig. 4) indicate inhomogeneous thickness of the network, while the highest detected points (~ 400 nm) are linked to lifted flakes extremities, the flakes which are localized on the top but not at the edges of patterns (see also fig. 3, HR images). The flakes at network edges making interface with glass lie flat confirming significant adhesion interactions (thickness of few nanometers is detected at the edges, fig.4 right, see also fig.3 HR images), while the top curved flakes suggests an excess of the flakes, which is linked to excessive initial concentration.

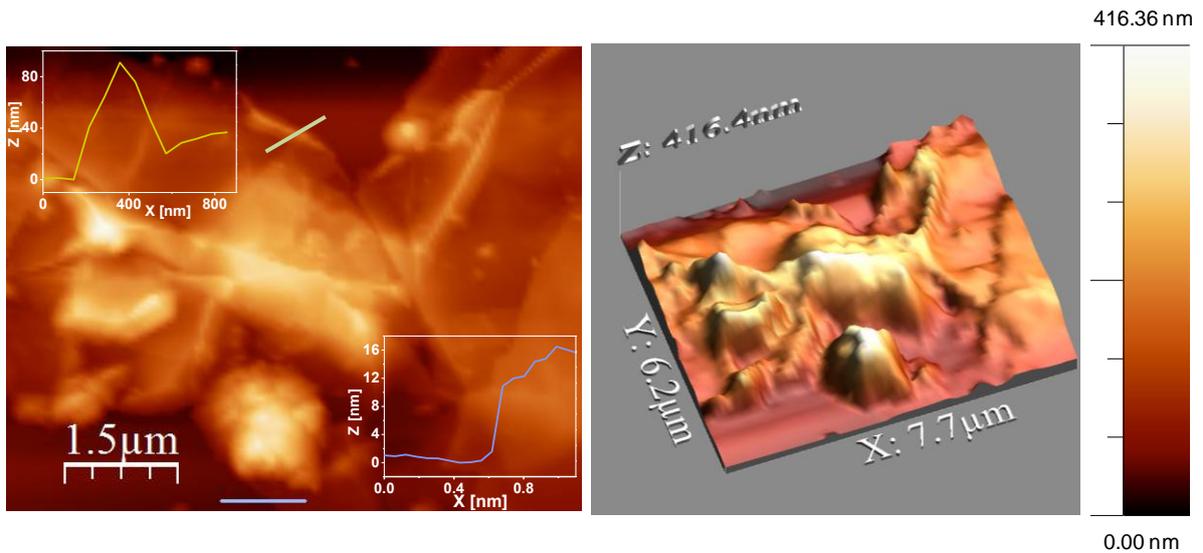

Figure 4. The AFM image with two profiles performed at the edges of patterns (right), and corresponding AFM tomography profile.

The fractal dimension ($D_f$) can be defined as $\log N(l)/\log l$, where "$N(l)$" is a number of boxes and "$l$" is a length of the boxes. $D_f$ was estimated for f-3 and f-4 by box counting method via linear fitting of the logarithmic relation with the "ImageJ" software.[42] Prior to the counting the chosen SEM micrographs were converted into binary images, which are next divided by squares (boxes) with varied size from 2 to 64 pixels. Fig. 5 demonstrates the representative binary images of f-3 and f-4 with corresponding $\log N(l)/\log l$ plots. The $D_f$, a slope of the plots are $1.73 \pm 0.02$ for both networks and suggests diffusion limited aggregation mechanism (DLA).[27,28] For few images $D_f$ goes up to higher value, $1.85 \pm 0.02$, which can be related to cross-linking between the branches.

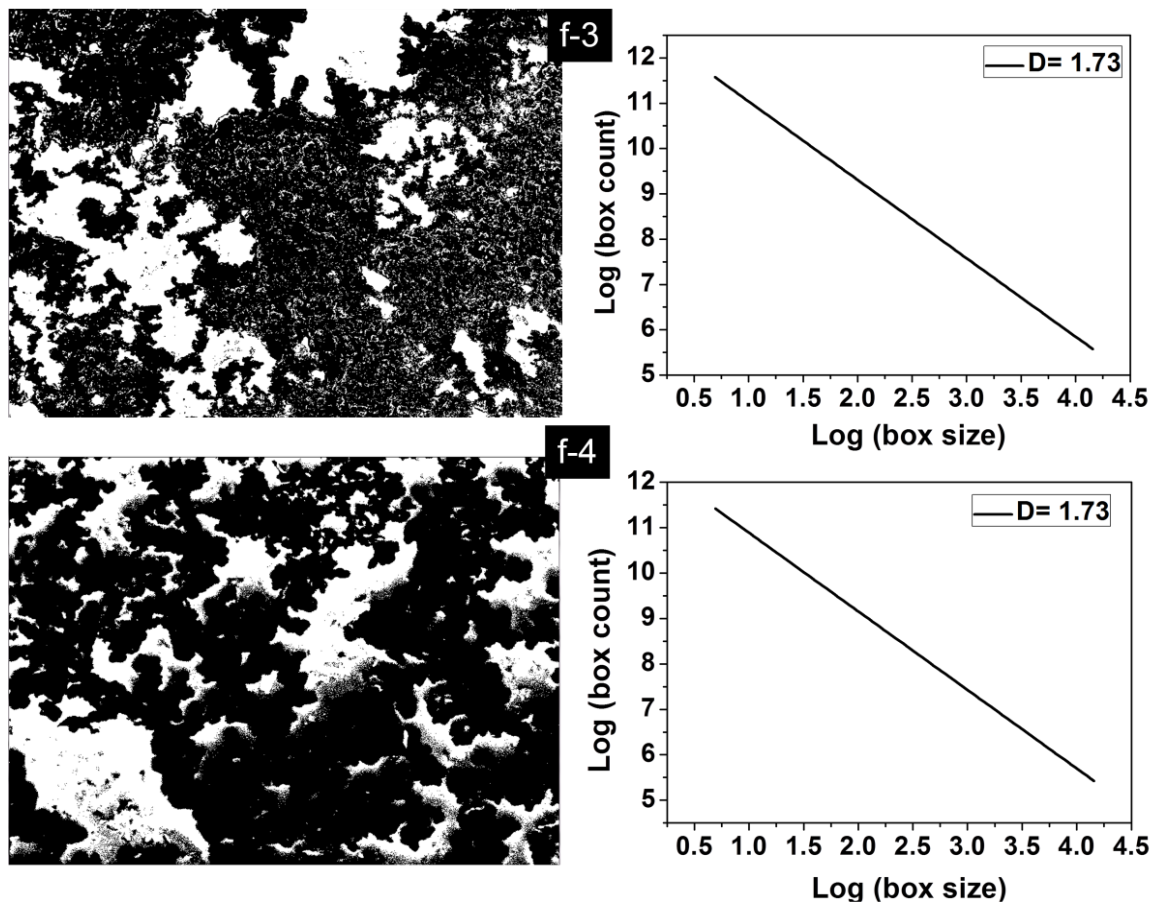

Figure 5. Binary images of f-3 and f-4 and corresponding logN(l)/logl plots.

The charge transport properties of f-1, f-3 and f-4 networks were investigated by four point probe (FPP) method with distance fixed probes, by injection of current in the range up to $10^{-2}$A. The experimental values were multiplied by geometrical correction factor according to the method protocol of FPP method for thin flms.[43] The measurements in addition were performed after annealing of the samples at 400°C in He for 4h, which aimed to removed adsorbed impurities and solvents residues. The networks exhibit ohmic behaviors; representative I(V) curves with corresponding calculated resistances are presented in fig.6. The resistance values before annealing are 15.4 kΩ, 2.4 kΩ for f-1 and f-3 respectively (high density center omitted), and 4.4 kΩ and 541 Ω for f-4 (at lower and higher density). The resistance values significantly drop after annealing treatment an order of magnitude to 2.3 kΩ, 130 Ω for f-1 and f-3 respectively and to 97 Ω and 48 Ω for f-4. The relatively lower resistance of f-4 (48 Ω, 97 Ω depending on local thickness) compared to FLG-Abl are in agreement with higher conductivity of individual FLG-EG components, however other factors such as flakes identity, different evaporation conditions and related pattern morphology can play an additional role.

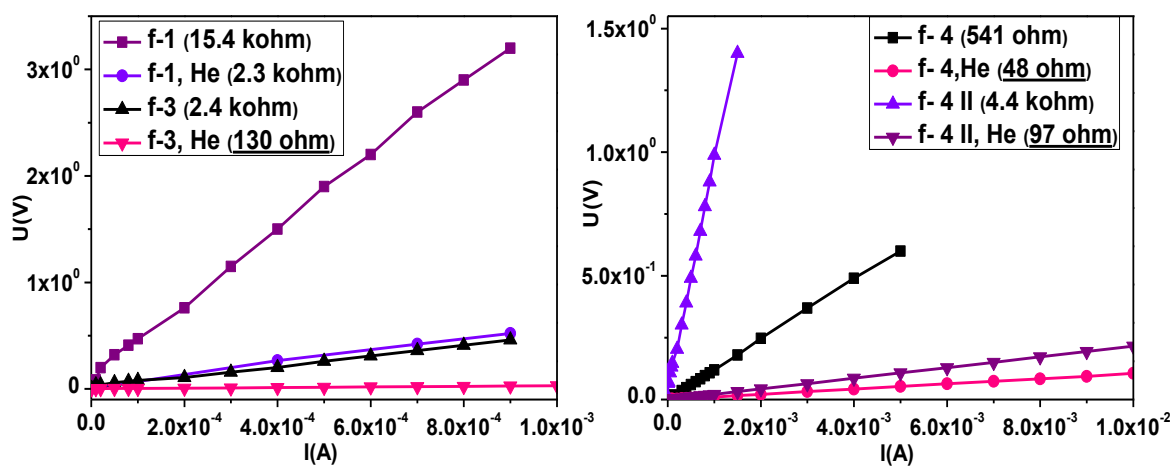

Figure 6. Representative I(V) curves obtained for f-1, f-3 and f-4 by four point method with corresponding resistance values.

The main finding from the charge transport measurements is that the percolation threshold is strongly reduced in the branched, fractal-like networks compared to a random arrangement, which confirms preliminary simulation results for self-assembled gold nanoparticles.[36] The performed earlier investigations considering FLG-Abl flakes shows, that when they are randomly deposited on glass substrate by hot spray techniques, a charge transport performance is much lower. The minimal resistance of the formed film and at minimal coverage, after annealing at 800°-900°C, was 760 Ω by Hall Effect technique, while the FPP measurements showed 85 kΩ and 15 kΩ before and after annealing respectively. Here, obtained by the FPP Rs for assemblies are one, two orders of magnitude lower, moreover the conductive film with random arrangement could not be achieved for a transparency higher than 32% (high coverage).[44] It is however difficult to determine the exact impact degree of self-assembly on conductivity improvement due to the undetermined graphene-graphene junction and local thickness variation. One thing is certain; the presented self assembly allows to occur a percolation at lower FLG concentration for a given surface, while the low coverage of substrate surface can increase additionally a transparency of the film. To improve a transparency initial concentration of the flakes needs to be also much lowered to avoid excessive stacking in "z" direction.

High conductivity and related continuity of the fractal networks are in addition reflected by SEM micrographs quality. A very low charging effect of the assemblies occurs during the analysis despite the large and thick insulator substrates of which significant surface area is FLG free. Beside the observed black FLG conductive paths, separated and charged FLG flakes can be punctually identified on network-free areas (flashes, fig.2 and 3).

The ohmic behavior of the fractal networks and such low sheet resistance confirm first: a high quality of the flakes (diffusive motion of electrons is limited only by scattering at functional groups and lattice defects);[45] second: an overlapping of the flakes (HR-SEM, fig.2 and 3) and formation of bilayer boundary region via van der Waals forces, which allows electron flow to overcome edges defect by connection of $sp^2$ carbons from overlapped flakes (contrary to direct atomic bonding with discrete atomic domain boundary),[46] third and principal: the self-assembled branched network strongly helps the percolation to occur compared to random assembly. The latter is of huge importance in view of (opto)electronic applications, where continuous branched networks constitute a path for charge transport, and when assisted by low coverage of substrate increases a transparency of the film similarly to graphene-carbon nanotubes hybrids or graphene mesh approach.[12-15] Moreover, according to "open" domain simulations branched structures can be reached at low surface coverage, only 10%.[26]

In conclusion, it should be underline that the presented evaporation conditions were not optimized with respect to (transparent) conductive films, and mainly aimed to demonstrate a possibility to form macroscopic self-assembled conductive graphene networks from (micro)nanoscopic matter with reduced percolation threshold (lower matter concentration is sufficient to reach the percolation for a given surface compare to random arrangement), via easy, cost-effective and substrate-size unlimited method. Several parameters need to be set in order to fully understand the graphene fractals-like self-assembly formation and to control conductivity and (or) transparency property in most important manner. The thickness of the networks can be optimized by the use of lower initial concentration and subsequent higher dispersion of flakes containing lower sheets number; and adapted evaporation conditions. The appropriate choice of the conditions would modify a conductivity/transparency ratio withtin a given matter.

The above results confirm the importance of recent trends dealing with implementation of bio-inspired architectures in materials science.

**Acknowledgements**

I would like to thank "Conectus" program (Alsace) for financial support; Yue Feng Liu and Azhar A. Pirzado for help during SEM microscopy analysis and four point probe measurements.

**References**

1. S. Bae, H. Kim, Y. Lee, X. Xu, J. S. Park,Y. Zheng, J. Balakrishnan et al. *Nat Nanotechnol* 2010, **5**,574–578.

2. N. Rouhi, Y. Y. Wang, P. J. Burke, *Appl. Phys. Lett.* 2012, **101**, 263101−1−263101−3.

3. A. Raina, X. Jia, J. Ho, D. Neizch, H. Son, V. Boulvic, M. S. Dresselhaus, J. Kong, *Nano Lett.* 2009, **9 (1)**, 30−35.

4. K. S. Kim, Y. Zhao, H. Jang, S. Y. Lee, J. M. Kim, J. H. Ahn, P. Kim,J. Y. Choi, B. H. Hong, *Nature*, 2009, **457**, 706–710.

5. P. Y. Huang, C. S. Ruiz-Vargas, A. M. van der Zande, W.S. Whitney, M. P. Levendorf, J. W. Kevek, S.Garg, J. S. Alden, C. J. Hustedt, Y. Zhu, J. Park, P. L. McEuen, D. A. Muller, *Nature* 2011, **469**, 389–392.

6. M. Cai, D. Thorpe, D. H. Adamson, H. C. Schniepp, *J.Mater.Chem.* 2012, **22**, 24992–25002.

7. S. Park, R. S. Ruoff, *Nature Nanotech.* 2009, **4**, 217 – 224.

8. C. K. Chua, M. Pumera, *Chem. Soc. Rev.* 2014, **43**, 291-312.

9. V. H. Pham, T. V. Cuong, S. H. Hur, E. W. Shin, J. S. Kim, J. S. Chung, E. J. Kim, *Carbon* 2010, **48(7)**, 1945–1951.

10. K. Rana, J. Singh, J.-Hyun Ahn, *J. Mater. Chem. C* 2014, **2**, 2646-2656.

11. D. Arthur, R. P. Silvy, P. Wallis, Y. Tan, J.-D. R. Rocha, D. Resasco, R. Praino, W. Hurley, *MRS Bulletin* 2012, **37**, 1297-1306.

12. V. C. Tung, L-M. Chen, M. J. Allen, J. K. Wassei, K. Nelson, R. B. Kaner, Y. Yang, *Nano Lett.* 2009, **9 (5)**, 1949–1955.

13. R. Akilimali, N. Macher, A. Bonnefont, D. Bégin, I. Janowska, C. Pham-Huu, *Materials Letters* 2013, **96,** 57-59.

14. S. H. Kim, W. Song, M. W. Jung, M.-A. Kang, K. Kim, S.-J. Chang, S. S. Lee, J. Lim, J. Hwang, S. Myung, K.-S. An, *Adv. Mater.* 2014, **26**, 4247–4252.

15. Q. Zhang, X. Wan, F.Xing, L. Huang, G. Long, N. Yi, W. Ni, Z. Liu, J. Tian, Y. Chen, *Nano Research* 2013, **6(7)**, 478–484.

16. B. B. Mandelbrot, The Fractal Geometry of Nature, W.H. Freeman, New York, 1983.

17. S. S. Manna, B. Subramanian, *Phys.Rev. Lett.* 1996, **76**, 3460-3463.

18. S. Vrobel, Studies of Nonlinear Phenomena in Life Science, vol.14., Fractal Time: Why a Watched Kettle Never Boils, World Scientific Publishing Co.Pte.Ltd., Singapour, 2011.

19. E. E. Peters, Fractal Market Analysis: Applying Chaos Theory to Investment and Economics, Johnson Wiley & Sons, Inc. 1994.

20. I. Vancea, U. Thiele, *Phys. Rev. E* 2008, **78**, 041601-1 - 041601-15.

21. E. Pauliac-Vaujour, A. Stannard, C. P. Martin, M. O. Blunt, I. Notingher, P. J. Moriarty, I. Vancea ,U. Thiele, *Phys. Rev.Lett.* 2008, **100**, 176102-1 - 176102-4.

22. K. Mougin, H. Haidara, *Langmuir* 2002, **18**, 9566-9569.

23. R. D. Deegan, O. Bakajin, T. F. Dupont, G. Huber, S. R. Nagel, T. A. Witten, *Nature* 1997, **389**, 827.

24. B. J. Fischer, *Langmuir* 2002, **18**, 60-67.

25. E. Rabani, D. R. Reichman, P. L. Geissler, L. E. Brus, *Nature* 2003, **426**, 271-274.

26. A. Crivoi, F. Duan, *Phys.Chem.Chem.Phys.* 2012, **14**, 1449-1454.

27. T.A. Witten, Jr., L. M. Sander, *Phys.Rv, Lett.* 1981, **47(19)**, 1400-1403.

28. T.A. Witten, L.M. Sander, *Phys.Rev.B.* 1983, **27 (9)**, 5686-5696.

29. P. Meakin, *Phys. Rev.A* 1983, **27(3)**, 1495-1507.

30. H.G.E. Hentschel, *The structure and fractal dimension of cluster-cluster aggregates 117-120 in Kinetics of Aggregation and Gelation.* Elsevier Science Publishers B.V., 1984.

31. W. D.Schaefer, J. E. Martin, *Aggregation of colloidal silica in Kinetics of Aggregation and Gelation.* 71-74. Elsevier Science Publishers B.V.; 1984.

32. S. Nie, J. M. Wofford, N. C. Bartelt, O. D. Dubon, K. F. McCarty, *Phys. Rev. B* 2011, **84**, 155425 –1- 155425 –7.

33. J. M. Wofford, E. Starodub, A. L. Walter, S. Nie, A. Bostwick, N. C Bartelt, K. Thürmer, E. Rotenberg, K. F. McCarty, O. D. Dubon, *New.J.Phys*. 2012, **14**, 053008.

34. X. Li, C. W. Magnuson, A.Venugopal, R. M. Tromp, J. B. Hannon, E. M. Vogel, L. Colombo, R. S. Ruoff, *J. Am.Chem.Soc.* 2011, **133**, 2816–2819.

35. D. Geng, B. Wu, Y. Guo, B. Luo, Y. Xue, J. Chen, G. Yu, Y. Liu, *J. Am. Chem. Soc.* 2013, **135 (17)**, 6431–6434.

36. D. G. Reuven, H. B. Mihiri Shashikala, S. Mandal, M. N. V. Williams, J. Chaudhary, X.-Q. Wang, *J. Mater. Chem. B* 2013, **1**, 3926-3931.

37. D. G. Reuven, K. Suggs, M. D. Williams, X.-Q. Wang, *ACS Nano* 2012, 6(2), 1011-1017.

38. Ching-Ling Hsu, Szu-Ming Chu, Kiwi Wood, Yi-Rong Yang, *Phys. Stat. Sol. (a)* 2007, **204( 6)**, 1856–1862.

39. C.-L. Hsu,Y.-R. Yang, K. Wood, *Chinese Journal of Phys.* 2007, **45 (6-II)**, 686-692.

40. I. Janowska, T. Romero, P. Bernhardt, F. Vigneron, D. Begin, O. Ersen, M.-J. Ledoux, C. Pham-Huu, *Carbon* 2012, **50**, 3092-3116.

41. M. S. Moldovan, H. Bulou, Y. J. Dappe, I. Janowska, D. Bégin, C. Pham-Huu, O. Ersen *J. Phys. Chem. C* 2012, **116 (16)**, 9274–9282.

42. B. B. Mandelbrot, *Les objects fractals: forme, hazard et dimension*, Flammarion, Paris, 1975. 43. F. M. Smit, *The Bell System Technical Journal*, May 1958, p. 711.

44. A. A. Pirzado, Y. Jouane, F. Le Normand, R. Akilimali, V. Papaefthimiou, C. Matei Ghimbeu, I. Janowska, *J. Phys. Chem. C*, 2014, **118 (2)**, 873–880.

45. C. Punckt, F. Muckel, S. Wolff, I. A. Aksay, C. A. Chavarin, G. Bacher, W. Mertin, *Appl. Phys. Lett.* 2013, **102**, 023114-1− 023114-5.

46. A. W. Robertson, A. Bachmatiuk, Y. A. Wu, F. Schaffel, B. Rellinghaus, B. Buchner, M. H. Rummeli, J. H. Warner, *ACS Nano* 2011, **5(8)**, 6610 − 6618.